%% file: coling-main.tex
\documentclass[10pt, a4paper]{article}

\usepackage[review]{lrec-coling2024} % this is the new style
\pdfoutput=1

\usepackage{xcolor}
\usepackage{hyperref}
\usepackage{natbib}
\usepackage{multibib}
\usepackage{amsmath,amssymb,amsfonts}
\usepackage{graphicx}
\usepackage{tikz}
\usepackage{pgfplots}
\usepackage{tabularx}
\usepackage{soul}
\usepackage{titlesec}
\usepackage{xstring}
\usepackage{booktabs}
\usepackage{colortbl}
\usepackage{bm}
\usepackage{multirow}
\usepackage{changes}
\usepackage{algorithmic}
\usepackage[ruled,linesnumbered]{algorithm2e}
\SetKwInOut{Input}{\textbf{Input}}
\SetKwInOut{Output}{\textbf{Output}}
\SetKwData{State}{\textbf{State: }}

\usepackage{color}

\usepackage{url}
\usepackage{makecell}

\usepackage{xspace}
\usepackage{color}
\definecolor{dark_red}{rgb}{0.5, 0, 0}
\definecolor{red}{rgb}{.9,0,0}
\definecolor{dark_green}{rgb}{0, 0.5, 0}

\newcommand{\mj}[1]{{\color{black}#1}}

\usepackage{xspace}
\makeatletter
\DeclareRobustCommand\onedot{\futurelet\@let@token\@onedot}
\def\@onedot{\ifx\@let@token.\else.\null\fi\xspace}
\def\eg{\emph{e.g}\onedot}

\newcommand{\Mat}[1]{\bm{#1}}
\newcommand{\Vector}[1]{\bm{#1}}

\newcommand{\Loss}{\mathcal{L}}

\title{\mj{Prior-agnostic} Multi-scale Contrastive Text-Audio Pre-training \\for Parallelized TTS Frontend Modeling 
}

\name{Author1, Author2, Author3} 

\address{Affiliation1, Affiliation2, Affiliation3 \\
         Address1, Address2, Address3 \\
         author1@xxx.yy, author2@zzz.edu, author3@hhh.com\\
         \{author1, author5, author9\}@abc.org\\}

\input{./sections/1_abstract.tex}
\begin{document}
\maketitleabstract
\input{./sections/2_introduction.tex}
\input{./sections/3_related_work.tex}
\input{./sections/5_method.tex}
\input{./sections/6_experiments.tex}
\input{./sections/7_conclusion.tex}

\section{References}
\bibliographystyle{lrec-coling2024-natbib}
\bibliography{reference}

\end{document}

%% file: sections/1_abstract.tex
\abstract{
Over the past decade, a series of unflagging efforts have been dedicated to developing highly expressive and controllable text-to-speech (TTS) systems. In general, the holistic TTS comprises two interconnected components: the frontend module and the backend module. The frontend excels in capturing linguistic representations from the raw text input, while the backend module converts linguistic cues to speech. The research community has shown growing interest in the study of the frontend component, recognizing its pivotal role in text-to-speech systems, including Text Normalization (TN), Prosody Boundary Prediction (PBP), and Polyphone Disambiguation (PD). Nonetheless, the limitations posed by insufficient annotated textual data and the reliance on homogeneous text signals significantly undermine the effectiveness of its supervised learning. To evade this obstacle, a novel two-stage TTS frontend prediction pipeline, named TAP-FM, is proposed in this paper. Specifically, during the first learning phase, we present a Multi-scale Contrastive Text-audio Pre-training protocol (MC-TAP), which hammers at acquiring richer insights via multi-granularity contrastive pre-training in an unsupervised manner. Instead of mining homogeneous features in prior pre-training approaches, our framework demonstrates the ability to delve deep into both global and local text-audio semantic and acoustic representations. Furthermore, a parallelized TTS frontend model is delicately devised to execute TN, PD, and PBP prediction tasks, respectively in the second stage. Finally, extensive experiments illustrate the superiority of our proposed method, achieving state-of-the-art performance.
\\ \newline \Keywords{Frontend modeling, Text-to-speech, Contrastive Text-audio Pre-training, Multi-modal learning}}

%% file: sections/2_introduction.tex
% 写作思路，
%   1、TTS系统有两种类型，1）端到端系统；2）前端后端分离的系统。而我们的研究对象是后者的前端模型部分。前后端分离的系统相对端到端来说具有的优势，以及前端模型的意义。
%   2、介绍前端模型的组成
%       1、串行模型
%       4、联合模型
%       5、总结不足之处，介绍最新的工作进展，CLAP，大模型
%   3、介绍我们的工作的动机创新和贡献

\section{Introduction}

Currently, text-to-speech (TTS) technology has exhibited its versatile applicability in a wide range of practical scenarios, such as news broadcasting, FM radio, digital personas, and intelligent voice assistants \cite{terzopoulos2020voice}. TTS aims to implicitly pursue the stable acquisition of phonemes from words, ultimately transforming raw text input into the desired audio output. \mj{Albeit the impressive performance, end-to-end {setting} of TTS~\cite{wang2017tacotron,ren2020fastspeech} heavily hinges on the {availability} of substantial {corpora} to {generate high-fidelity pronunciation}. {Starting with this limitation, we divert} attention to the separated systems involving the frontend and backend modules thanks to their {accuracy}, controllable and customizable speech synthesis outputs. Wherein,} the frontend focuses on extracting linguistic cues from \mj{text inputs while} the backend part {is responsible for converting linguistic representations} into natural speech.

% 【todo】add引用
Typically, the frontend in an English TTS operates in a serial-pipelined fashion, which is comprised of Text Normalization (TN), Prosody Boundary Prediction (PBP), and Grapheme-to-Phoneme (G2P). {Concretely, 
\emph{i)}} TN component takes the texts as input and categorizes non-standard words (NSWs) into {varied} labels (\eg money, time, and date). {Label-specific rule-based functions are then utilized} to verbalize these NSWs into spoken-form words (SFWs);
{\emph{ii)} PBP attempts to} predict three types of labels: {\emph{Prosodic Word (PW), Prosodic Phrase (PPH) and Intonational Phrase (IPH)}}, whereas {\emph{iii)}} G2P is ultimately devoted to transforming words into phoneme sequences. The {entire} G2P procedure consists of three essential components: \emph{i)} Polyphone Disambiguation (PD) {estimates correct phone for each polyphone through a classification behaviour} estimates correct phone for each polyphone; 
\emph{ii)} Pronunciation Dictionary Matching (PDM) aims to look up phoneme sequences for words in dictionary {which is comprised of} word-phoneme pairs; \emph{iii)} Letter-to-sound (LTS) is in charge of predicting phonemes for out-of-vocabulary words. {As heuristically-designed PDM and LTS are context-independent and impervious to capacities of diverse models, we particularly delve deep into behaviour of PD in this paper.}
While a pipeline-based system provides greater flexibility and controllability by breaking down complex tasks into smaller steps, the demands for extra training procedures, storage, and computational {overheads easily} become formidable. In addition, the noises introduced {from} upstream tasks are prone to severely {degrade} the final performance. {Therefore, these} drawbacks spur the booming of unified models~\cite{unified-mul, conkie2020scalable, pan2020unified, bai2021universal, ying2023unified} which integrate all tasks in a unified framework. In spite of the attractive performance, these existing approaches still {partially/fully} inherit the {unsatisfactory} properties of the pipeline-based structures, thereby hindering the development of paralleled protocol. Deep parallelism {allows to simultaneously process} all frontend modules via a unified language model. {Considering} the richer semantic priors offered by large-scale pre-trained models (\eg BERT~\cite{BERT}), both current pipeline- and unified-based systems~\cite{zhang2020hybrid, Span-PB,dai2019disambiguation,bai2021universal} utilize these pre-trained models to advance learning capabilities of their approaches through unsupervised learning on extensive {unlabled} text data, reducing the needs for time-consuming and labor-intensive annotations for {multifarious} frontend tasks. {However, these methods always produce unsatisfactory diversity of features and information (\eg intonation, emotion and speaker characteristics)}, mainly resulting from overlooking multi-modal data (audio).

To address this issue, SpeechT5 \cite{ao2021speecht5} makes pioneering {efforts} to learn representations from text-audio pairs via a unified multi-modal pre-training framework. CLAPSpeech \cite{clapspeech} incorporates {a host of} text-phone pairs to further enhance {TTS performance}. {Even so}, CLAPSpeech is incapable of capturing globally-oriented contextual cues, and {requires auxiliary tools to manually achieve the alignment between text and audio data.} Therefore, there is still significant room for improvement in enriching representations in TTS systems.

{To mitigate these drawbacks, in this paper, we delicately devise a two-stage pipeline, namely TAP-FM. The first stage endeavors to enrich text-audio semantics and acoustic hints via a novel Multi-scale Contrastive Text-Audio Pre-training framework (MC-TAP) consisting of three well-designed pre-training tasks. They are}: \emph{(1)} \emph{Span-level} contrastive learning for text-audio pairs aims at capturing local dependencies in the audio context. {Instead of using data preprocessing to align text \& audio segments in previous works, we propose a novel fully-automatic labeling alignment mechanism by resorting to the dynamic programming algorithm;} \emph{(2)} \emph{Sentence-level} contrastive learning {endows the model with higher discriminability among different speakers and stronger generalizability;} \emph{(3)} \emph{Masked Language Modeling (MLM)} {is conducive to preserve original semantic information when the model is undergoing the learning of} acoustic characteristics. {To deal with TN, PBP, and PD tasks in parallel, we establish a framework based on} the TextEncoder {in} MC-TAP and {a set of manually-defined rules in the second stage}. {Specifically, during the course of joint fine-tuning of multiple tasks, the improved Dynamic Weight Averaging (DWA)~\cite{dwa} is introduced to suppress the negative effects brought from distinct data sources for diverse tasks. In addition, to inherit the merits of residual learning, the residual connection is also designed for Conformer~\cite{gulati2020conformer} to enhance the model performance while accelerating convergence.}

In a nutshell, the contributions in this paper are threefold:
\begin{itemize}
    \item {\bf Fine-grained Representations.} {In this paper, a  prior-agnostic framework, namely MC-TAP, is novelly proposed to carry out multi-scale contrastive pre-training, thereby learning fine-grained representations including both sentence-level (global) and span-level (local)} text-audio semantics and acoustic representations.
    \item {\bf Paralleled {Paradigm}.} {To gain satisfactory efficiency and performance, this paper presents a highly efficient \& parallelized model for TTS frontend prediction on top of the text encoder of the MC-TAP approach. Moreover, a modified DWA technique is adopted to} address the issue of imbalanced and non-uniform data across the multiple subtasks.
    \item {\bf SOTA Performance.} {Extensive} experiments demonstrate the superiority of our proposed method, {leading to state-of-the-art (SOTA) performance on TTS frontend multi-subtask predictions. Furthermore, necessary studies are also carried out to ablate our proposed individual components.}
\end{itemize}

%% file: sections/3_related_work.tex
\section{Related Work}
{In this section, we review recent related works which are roughly categorized into three types:} 1) Pipeline-based Frontend Modeling, 2) Unified Frontend Modeling, and 3) Multi-modal Contrastive Pre-training Modeling.

\subsection{Pipeline-based Frontend Modeling}
%We review prior works including TN, PBP, and PD in pipeline-based frontend modeling. 

\textbf{Text Normalization.} The early {rule-based TN task ~\cite{zhou2008document,ebden2015kestrel} allows for controlling,} but makes it hard to adjust rule priorities, solve ambiguity, and resolve conflicts among rules. Thanks to the huge success of deep learning~\cite{he2016deep,han2019image,han2022dual}, machine learning-based approaches \cite{crf-tn, sproat2016rnn, lusetti2018encoder, 9415113} {is capable of disambiguating}
the same words in different contexts and achieve higher overall accuracy. {However, they} sometimes produce unreasonable results due to the lack of constraints. {To tackle this issue}, hybrid methods \cite{jiang-etal-2020-cold,gokcen2019dual,zhang2020hybrid,flattn} {catch growing attention from the research community thanks. They consider both rules and neural networks, yielding higher accuracy and explainability.}

\textbf{Prosody Boundary Prediction.}
The rule-based method \cite{zhang2016mandarin} utilizes syntactic trees to predict prosodic phrases.
{The work in}~\citet{qian2010automatic} predicts prosody using CRF model, {whereas \citet{blstmPB} uses {BiLSTM and CRF to encode and decode representations, respectively.} A framework \cite{pan19b_interspeech} that combines BiLSTM-CRF and Multi-Task Learning (MTL) {are} proposed to deal with the issue. Pre-trained language models that leverage unsupervised data have been applied as well in SpanPSP \cite{Span-PB}, which utilizes a pre-trained text encoder and a span-based text feature extraction method to yield better results. Study \cite{lee2023cross} on PBP cross-lingual transfer {illustrates that a pre-trained multilingual model contributes to enhancing zero- or few-shot transferability.}

\textbf{Polyphone Disambiguation.} Polyphone is common in many languages such as English and Mandarin. The pronunciation of synthesized audio is directly affected by the prediction of polyphones. Study \cite{mao2007inequality} has used maximum entropy (ME) to classify each polyphone, and a unified method \cite{gorman2018improving} has proposed to incorporate rules and machine learning. More recent researches \cite{nicolis2021homograph,sun2019knowledge,dai2019disambiguation,zhang2022polyphone, hida2022polyphone,zhang2021polyphone} have focused on applying pre-trained language models to PD task.
% huang2010disambiguation【规则+unified】是否加入【todo】

\subsection{Unified Frontend Modeling}
{Due to the fact that} pipeline-based frontend modeling is costly and prone to error accumulation, {researchers gradually pay more attention to unified frontend modeling.} ~\citet{unified-mul} adopts knowledge distillation {technique} to compress the BERT model and achieves joint training for the PBP and PD tasks.
Other works \cite{conkie2020scalable, pan2020unified, bai2021universal} aim to improve unified models as well. A newer research, UFEF \cite{ying2023unified}, further integrates rule-based methods on BERT, combining the tasks of TN, PBP, and G2P and sharing model parameters during training.

\subsection{\fontsize{10pt}{12pt}\selectfont Multi-modal Contrastive Pre-training Modeling}
In recent years, the TTS system has rapidly incorporated multi-modal approach. Inspired by contrastive learning of text-image \cite{clip}, CLAP \cite{elizalde2023clap,wu2023large} develops a contrastive language-audio pre-training model. CLAPSpeech \cite{ye2023clapspeech} then utilizes CLAP to learn hidden features of prosody in audios, thereby improving prosody for existing TTS systems. Another application of CLAP \cite{SSWP} involves annotating prosody boundaries based on text-audio pairs. However, these methods heavily rely on forced alignment tools to align texts and audio, which require additional data processing. And incorrect alignment can negatively impact the final results. Additionally, a bootstrapped Seq2Seq frontend \cite{sun2023improving} that utilizes audios has been proposed. It is distilled from an existing pipeline-based frontend and updated using phoneme sequences decoded from transcribed speech audio.
% Yet this method is subjected to no restriction and no audio information during training and inferring.
% 但是这种方法在训练和推理时并没有真正的引入音频信息。
However, this method does not truly introduce audio information during training and inference.

%% file: sections/5_method.tex
\section{Methodology}\label{sec:model}
\begin{figure*}
\centering
\includegraphics[width=0.95\linewidth, trim=10 0 0 0, clip]{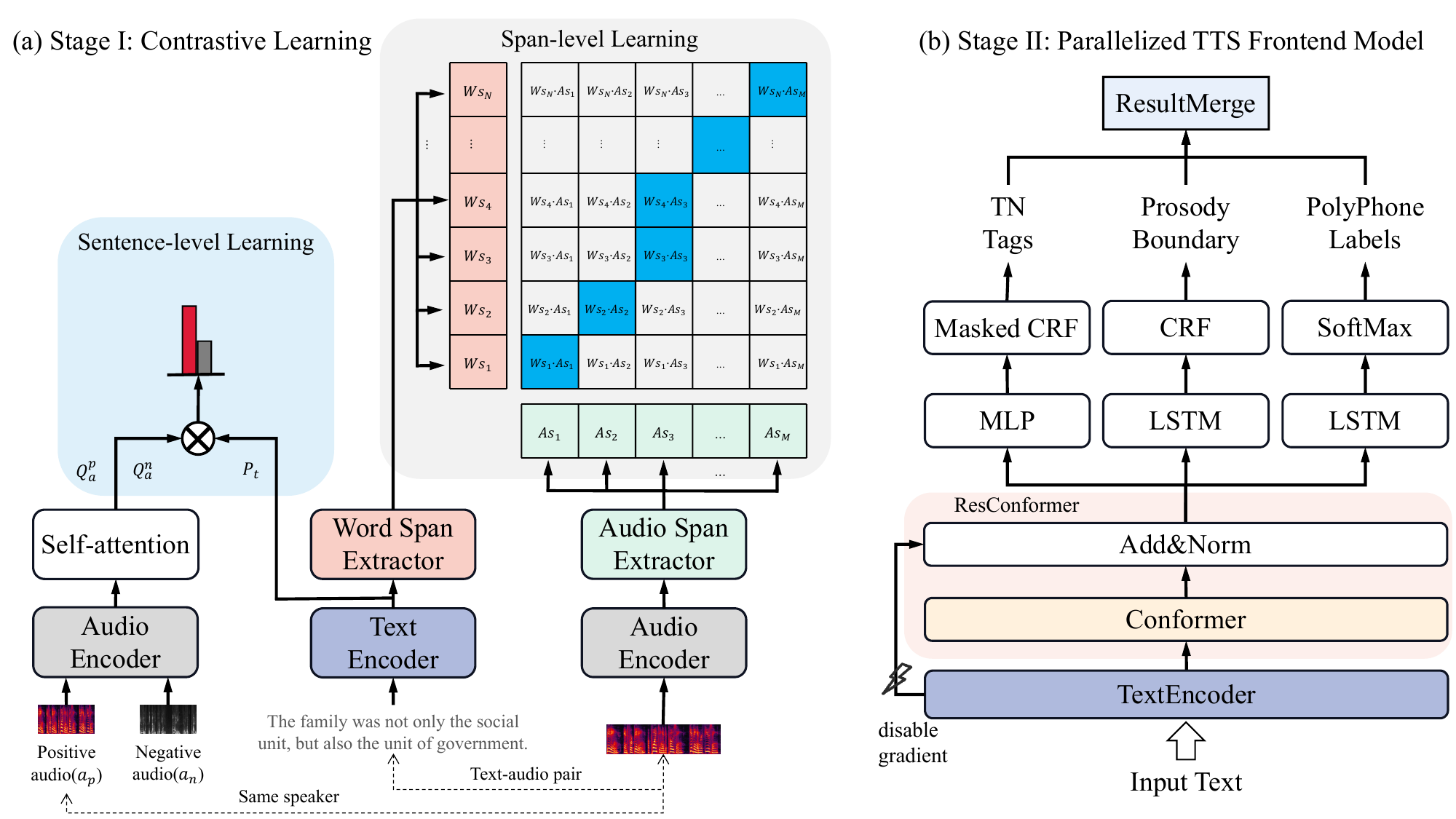}
\caption{Overview of the proposed system. Stage I: The prior-agnostic Multi-scale Contrastive Text-audio Pre-training (MC-TAP). Stage II:  Our parallelized TTS frontend model for predicting TN, PBP, and PD tasks. ResultMerge is a rule-based approach designed to address prosody boundary and polyphone issues within NSWs.
}
\label{fig:pre-model}
\end{figure*}
In this section, we basically introduce the proposed system in detail.
First, an overview of the system is illustrated.
Subsequently, we clarify the procedures of the multi-scale contrastive text-audio pre-training and frontend multi-subtask predictions.
% Finally, we elaborate on the training and prediction of our framework.

\subsection{Overview}

In order to improve the accuracy of parallelized TTS frontend models, we introduce a novel framework, Contrastive \textbf{T}ext-\textbf{A}udio \textbf{P}re-training for Parallelized TTS \textbf{F}rontend \textbf{M}odeling (\textbf{TAP-FM}), designed to enhance the capabilities of frontend models by incorporating audio-related features through text-audio prior-agnostic contrastive learning.
Figure \ref{fig:pre-model} depicts the overall architecture of TAP-FM, which is a two-stage process. 
The first stage involves prior-agnostic \textbf{M}ulti-scale \textbf{C}ontrastive \textbf{T}ext-\textbf{A}udio \textbf{P}re-training (\textbf{MC-TAP}). In this stage, we use a text encoder and an audio encoder to generate sequential representations for input text and the corresponding audio. Additionally, we perform span-level and sentence-level contrastive learning that enables TAP-FM to capture comprehensive audio information. In order to learn semantic information at the same time, we equip MC-TAP with masked language modeling (MLM).
In the second stage, we design a parallelized frontend model to predict TN, PBP, and PD tasks. The model incorporates the pre-trained TextEncoder from MC-TAP to capture fine-grained semantic and acoustic cues. Notably, we use ResConformer to optimize the direction of gradient propagation. To expedite model convergence, we employ a modified DWA method to repair the imbalanced and non-uniformity of the multiple-task datasets.
Below we provide a detailed introduction to the key components of the system.

\subsection{Multi-scale Contrastive Text-Audio
Pre-training}\label{Pre-traing}

The existing TTS frontend system solely relies on textual data for training the frontend models. Consequently, frontend models are unable to leverage the rich information embedded in the target audio. To address this inherent limitation, we devise a Multi-scale Contrastive Text-audio Pre-training protocol (MC-TAP) as illustrated in Figure \ref{fig:pre-model}(a). This approach facilitates the learning of feature alignment between text and audio, encompassing both semantic and acoustic domains, to improve the performance of the frontend models.

During this stage, for the text input of MC-TAP, denoted as $t$, we employ the pre-trained language model BERT \cite{BERT} as the text encoder to acquire a comprehensive overall representation $P_t$ and a sequential representation $E_t$. Specifically, $P_t$ corresponds to the feature vector of [CLS] at the top layer of BERT, while $E_t$ represents a sequence of vector representations, with each corresponding to a word in $t$. 
For audio input, denoted as $a$, we adopt WavLM \cite{chen2022wavlm} as the audio encoder to take the audio’s representation $E_a$. To ensure consistency with the text dimension, we apply a simple transformation using an MLP. Here, $E_a$ represents the vector representations of each segment in audio a. In the WavLM model, the global representation is not indicated by a specific flag. Instead, an attention mechanism is utilized to acquire the overall audio representation, referred to as $Q_a$.

% \begin{equation}\label{eq:qa}
% Q_a = \mathrm{Attention}(\Mat{E}_a)
% \end{equation}

\textbf{Span-level Automatic Alignment Comparative Learning.}
The intonation during speech is primarily influenced by the immediate context rather than long-range dependencies, which is particularly evident in everyday conversations. For example, during a discussion, individuals frequently maintain a consistent intonation and tone, contributing to the overall coherence of the conversation. This linguistic coherence plays a crucial role in fostering mutual understanding and effective communication.
To capture the strong close-range dependency, we employ overlapping sliding windows to slice the feature sequences of both text and audio into span-form. Specifically, we have developed the TextSpan Extractor and AudioSpan Extractor, which are based on BiLSTM, to incorporate contextual information from text and audio, respectively. It is important to note that these extractors have a common structure but do not share model parameters.

% For each span, we take the last output of BiLSTM as the overall representation. The sequence features of the text and audio spans can be represented separately 

For each span slicing, we utilize the last step output of BiLSTM as the span representation. The sequence features of the text and audio spans can be separately represented
as $\Mat{S}_t \in \mathbb{R}^{N \times D}$ and $\Mat{S}_a \in \mathbb{R}^{M \times D}$, where $N$ is the length of text spans, and $M$ is the length of audio spans.
% 受CLIP的启发，一个N*M的Span级别的配对矩阵可以被计算出来
Inspired by CLIP \cite{clip}, a span-level pairing matrix of size $N \times M$ can be calculated as follows:
\begin{equation}
    logits  = \mathrm{Norm}( \Mat{S}_t) \cdot \mathrm{Norm}( \Mat{S}_a^T )\,,
\end{equation}
where $\mathrm{Norm}$ represents L2 normalization.

Our objective is to integrate text and audio characteristics at the span level by maximizing the cosine similarity of the text and audio span embedding of the real pairs while minimizing the cosine similarity of other pairs. However, the limitation of MC-TAP is the absence of supervising labels that associate specific text spans with corresponding audio spans. Thankfully, we can leverage the monotonic increasing alignment relationship between text subwords and audio segments to address this issue. We design Algorithm \ref{alg:get_label} to generate the Text-to-Audio (T2A) mapping labels $Label_T2A$ and Audio-to-Text (A2T) mapping labels $Label_A2T$.
Taking T2A as an example, prior to executing the algorithm, we get the index of the maximum value in the pairing matrix as the candidate labels, defined as follows:
\begin{equation}
    Label_{T2A}^{'}  = \mathrm{Argmax}( logits )\,.
\end{equation}
Next, we input the T2A candidate label sequence $Label_{T2A}^{'}$ and the maximum label value $M-1$, denoted as $v_{max}$, into Algorithm \ref{alg:get_label}. In this algorithm, we first use a dynamic programming algorithm to find the longest increasing subsequence (FindLIS) from the candidate labels $Label_{T2A}^{'}$ as the true label. If it does not satisfy the LIS case, the label sequence values are adjusted based on the context. 
We use the cross-entropy loss for T2A alignment, defined as follows:
\begin{equation}
    \Loss_{T2A}  = \mathrm{CELoss}(logits, \, Label_{T2A})\,. 
\end{equation}
By transposing the logits and similar processes, we can calculate the $Label_{A2T}$ and $\Loss_{A2T}$.

In conclusion, span-level contrastive learning loss is represented as:
\begin{equation}
    \Loss_{span}  = \frac{\Loss_{T2A} + \Loss_{A2T}}{2}\,.
\end{equation}

\begin{algorithm}[t]\label{alg:get_label}
\caption{Span Label Generation}
\SetAlgoNlRelativeSize{-3}
\KwIn{candidate labels $\Vector{Label^{'}}$,\\ maximum value of the label $v_{max}$}

\textbf{State: }{$last = 0$, \, $L = len(\Vector{Label^{'}})$, \, $rate =  \frac{L}{v_{max}}$, \, $\Vector{Label} = [0]*L$}

$\Vector{indexes}, \Vector{values} \gets \text{FindLIS}(\Vector{Label^{'}})$ 

% \tcp{\footnotesize \textit{find the longest increase sub-\\sequence's indexes and values \\ from candidate labels}} 

\For{$i$ \textbf{in} \textbf{range}(1, $L$)}{
    \If{$i \in \Vector{indexes}$}{
        $\Vector{Label}[i] = \Vector{Label^{'}}[i]$
    }
    \Else{
        $temp = (\min(last+1, v_{max}) + \frac{i}{rate})/2$
        
        \If{$temp \in \Vector{values}$ \textbf{or} $temp < last$}{
            \tcp{\textit{ignored label in loss}}
            $\Vector{Label}[i] = -1$  
        }
        \Else{
            $\Vector{Label}[i] = temp$
        }
    }
    $last = \max(\Vector{Label}[i],last)$
    
}
$\Vector{Label}[-1] = v_{max}$

\KwOut{$\Vector{Label}$}
\end{algorithm}

% 通过算法1，我们可以计算得到

\textbf{Sentence-level Comparative Learning.}
To gain a more extensive and diverse understanding of different speakers, we have designed a global contrastive learning task at the sentence level. Specifically, we sample both positive and negative audio examples given the text $t$ and paired audio $a$ from speaker $s$.  The positive example, denoted as $a_p$, is another audio clip from speaker $s$, while the negative example, denoted as $a_n$, is a randomly selected audio clip from a different speaker.  This approach allows us to effectively contrast and compare different speakers in order to enrich our knowledge and insights. The corresponding global representations of the positive and negative audio examples are denoted as $Q^p_a$ and $Q^n_a$.
% 句子级别的损失函数定义为：
The sentence-level loss function is defined as:
\begin{equation}
    \Loss_{sen} = 2-\frac{{P_t \cdot Q_a^p}}{{\|P_t\| \|Q_a^p\|}}+\frac{{P_t \cdot Q_a^n}}{{\|P_t\| \|Q_a^n\|}}\,.
\end{equation}

\textbf{Mask Language Model Pre-training.}
% 为了保留TextEncoder原始的语义信息，我们借鉴了BERT的掩码语言模型，以15%的概率用【Mask】替换掉文本中的单词，并预测原始的单词
To preserve the original semantic information of the text-encoder, we borrow the masked language model (MLM) from BERT. Specifically, we randomly replace 15\% of the words in the text with the special token [MASK] and then predict the original words
The loss of this part is represented as $\Loss_{mlm}$.

The final loss during the pre-training phase is represented as:
\begin{equation}
    \Loss = \Loss_{span} + \alpha\Loss_{sen} + \beta\Loss_{mlm}\,,
\end{equation}
where $\alpha$ and $\beta$ are hyperparameters for loss weighting.

\subsection{Frontend Multi-subtask Predictions}

\textbf{Parallelized TTS Frontend Modeling.}
We propose a parallelized TTS frontend model for TN, PBP, and PD tasks as illustrated in Figure \ref{fig:pre-model}(b).  To leverage the benefits of linguistic and acoustic multi-modal model information, this model borrows the contrastive pre-trained TextEncoder from MC-TAP.  Furthermore, we form a Conformer-based residual network (ResConformer) to optimize the direction of gradient propagation, and the gradient of the residual connection is disabled during training.
For predicting subtask TN,  masked CRF \cite{wei2021masked-crf} is employed to classify their type in a sentence, following the ResConformer. This approach ensures that the predicted results for subwords are categorized correctly, thus preventing any invalid results. In the case of PBP task, LSTM with CRF is employed to model contextual information. Furthermore, for the PD subtask, LSTM with softmax is utilized to predict the pronunciation of each polyphonic word.

% \qx{to do DWA+}
% 对于多任务联合训练来说，通常需要平衡各任务的收敛速度，受到DWA的启发，我们提出了一种名为DWA+的更加有效的的自适应加权方法。原始的DWA当某个任务率先趋于收敛时会增大该任务的Loss权重，这是不合理的，为此我们引入了收敛系数来做平衡。

\noindent\textbf{DWA+.}
To accelerate the convergence of the model, this study introduces a more efficient adaptive weighting technique known as DWA+ (modified Dynamic Weighted Averaging). This method is specifically designed to address the imbalanced and non-uniform attributes of the multiple-task datasets during the training phase. The initial definition of DWA is as follows:
\begin{equation}
    \lambda_k(l) = \frac{K\,exp(r_k(l-1)/T)}{\sum_i{exp(r_i(l-1)/T)}},
\end{equation}
\begin{equation}
        r_k(l-1) = \frac{\Loss_k(l-1)}{\Loss_k(l-2)},
\end{equation}
% l表示迭代次数
where $\lambda_k(l)$ is the loss weight, $l$ represents the number of iterations, $r_k(\cdot)$ represents the relative descending rate of loss reduction, $T$ is the temperature parameter, $K$ represents the number of tasks, and $\Loss_k(l)$ is the average loss in each iteration.

% As indicated in formulas, the loss rate $r_k$ of a specific task exhibits a gradual increase as it approaches convergence. In contrast, the loss rate of the other subtasks may not necessarily demonstrate such a trend. 
However, there is a problem that DWA will increase the loss weight of a task when it tends to converge, while other subtasks do not demonstrate such a trend. 
This discrepancy will lead to inadequate learning of the model. To tackle this issue, our proposed DWA+ method integrates a convergence coefficient $\epsilon_k$ to alleviate the loss in such scenarios. The calculation of $\epsilon_k$ is outlined as follows:

\begin{equation}
    \epsilon_k(l) = \frac{K\,exp(d_k(l-1)/T)}{\sum_i{exp(d_i(l-1)/T)}}  \,,
\end{equation}

\begin{equation}
     d_k(l-1) = \frac{h_k^m - h_k(l-1)}{h_k^m}\,,
\end{equation}
where $d_k(\cdot)$ represents the degree of relative convergence, $h^m$ is the theoretically best metric score. This paper uses the F1-score as the metric.

The final loss weight is defined as follows:
\begin{equation}
    w_k(l)=\frac{\lambda_k(l)+\epsilon_k(l)}{2}\,.
\end{equation}

% to do并行推理工作流
% \qx{to do \textbf{Parallel inference workflow}}
% 一般的TTS前端系统会按照TN，PD,PBP的顺序来完成

%% file: sections/6_experiments.tex
\section{Experiments}
\subsection{Dataset}
% 在预训练阶段，我们使用了两个英语ASR数据集训练MC-TAP模型，第一个是LibriSpeech，它包含了1000个说话人360小时的语音以及对应的文本，另一个是GigaSpeech，包括从有声读物、播客和YouTube上收集了约250个小时的转录音频，以及对应的人工转录文本。这两个数据集涵盖诵读和自发口语等一系列不同风格，以及艺术、科学、体育等多种主题。
In the pre-training phase, we train the MC-TAP model using two English ASR datasets. The first dataset consists of 360 hours of speech data from 1000 speakers, which is derived from LibriSpeech \cite{panayotov2015librispeech}. Each speech segment is accompanied by its corresponding textual representation. The second dataset is obtained from GigaSpeech \cite{chen2021gigaspeech}, which provides approximately 250 hours of transcribed audio. These datasets are carefully curated to cover a wide range of speech styles, including reading and spontaneous speech. Moreover, the topics covered in these datasets span across domains such as art, science, and sports.
% In the pre-training phase, we train the MC-TAP model using two English ASR datasets. The first dataset, LibriSpeech \cite{panayotov2015librispeech}, comprises 360 hours of speech from 1000 speakers, each with corresponding text. The second dataset, GigaSpeech \cite{chen2021gigaspeech}, offers approximately 250 hours of transcribed audio sourced from audiobooks, podcasts, and YouTube, along with manually transcribed text. These datasets encompass a variety of speech styles, including reading and spontaneous speech, as well as a wide range of topics spanning art, science, and sports.

% 在下游任务上，TN的数据集使用的是谷歌的开源数据，PBP数据集使用的是自己标注的LJ数据，PD数据是开源的多音字音库数据。
% On downstream tasks, we use the following datasets: \emph{(1)} TN dataset is released by Sproat and Jaitly \cite{sproat2016rnn}.
% This dataset is composed of Wikipedia texts. Each token in the sentence is categorized into 15 categories such as PLAIN, DATE, CARDINAL, VERBATIM, etc., of which PLAIN is non-NSW tag and the others are NSW tag. Each NSW token has its transliterated spoken form. \emph{(2)} The PBP dataset is an internal dataset, which is attained by labeling three prosody levels on text based on the corresponding audio. The text-audio pairs are from LJ Speech \cite{ljspeech17}, a publicly available TTS corpus consisting of 13100 text-audio pairs. \emph{(3)} PD dataset is from Gorman et al \cite{gorman2018improving}.
% This dataset consists of sentences containing polyphonic words with labels. Each polyphone has more than 2 labels, each corresponding to a unique pronunciation.
For downstream parallelized frontend model training, we utilize the following datasets: 
\emph{(1)} The TN dataset \cite{sproat2016rnn} consists of Wikipedia texts. Each token in the sentence is assigned to one of 15 NSW categories, such as PLAIN, DATE, CARDINAL, etc. 
% The PLAIN category represents non-NSW tags, while the other categories represent NSW tags. Each NSW token includes its transliterated spoken form.
\emph{(2)} The PBP dataset is sourced from LJSpeech \cite{ljspeech17}, which is a publicly available TTS corpus comprising 13,100 text-audio pairs. We label three prosody levels on the text according to the audio. \emph{(3)} The PD dataset \cite{gorman2018improving} consists of sentences containing polyphonic words with labels. Each polyphone in the dataset has more than two labels, each corresponding to a unique pronunciation.

\subsection{Experimental Settings}
% 在TextEncoder中，我们使用了BERT-medium的版本，含有8层的Transformer网络，对于音频编码器，为了获取丰富的音频信息使用了预训练好的WavLM-large版本的模型，并在训练过程中冻结了他的参数。我们在4个3090上进行预训练，初始学习率是6e-5,使用AdamW更新参数，

% We use BERT as the TextEncoder, and for the AudioEncoder, we use a pre-trained WavLM model to obtain rich audio information and freeze its parameters during the training process. 
% The sliding window size for the text span is 3, with a stride of 1, and the sliding window size for the audio is 33, with a stride of 11.
% The hyperparameters $\alpha$ and $\beta$ for the loss weights during the pre-training phase are both 0.5.
% The temperature parameter $T$ in DWA+ is set to 2.
% We perform training on 4 3090 GPUs, with an initial learning rate of 6e-5, using AdamW to update the parameters.
In our approach, we use BERT as the TextEncoder and a pre-trained WavLM model as the AudioEncoder. The parameters of WavLM are fixed during the training process. For the text span, the sliding window size is 3 with a stride of 1. As for the audio, a sliding window size of 33 with a stride of 11 is used. During the training of MC-TAP, the loss weights hyperparameters $\alpha$ and $\beta$ are both set to 0.5. Regarding DWA+ in frontend modeling, the temperature parameter $T$ is set to 2. The training is conducted on four NVIDIA GeForce RTX 3090 GPUs, with an initial learning rate of 6e-5, using AdamW for parameter updates.

\subsection{Baseline Models}
% We compare the proposed TAP-FM with the following
% advanced neural network (NN) based TTS frontend models proposed in recent years:
% \begin{itemize}
%     % 以预训练的BERT为文本编码器，在三个子任务上分别使用MLP作为预测层，
%     \item \textbf{BERT-base} \cite{BERT}: Use pre-trained BERT as the text encoder and use MLP as the prediction layer for three subtasks.

%     % 一个字符级别的通用预训练模型，可以完成TTS和ASR任务，我们把其中TTS任务的文本编码器部分放到下游任务中作为基线模型。
%     \item \textbf{SpeechT5} \cite{ao2021speecht5}: A character-level universal pre-training model that can complete TTS and ASR tasks. We put the text encoder part of the TTS task into the downstream task as a baseline model.

%     % 一个统一的前端框架来捕获英语TTS前端模块之间的依赖关系，以BERT为骨架，结合了规则的方法
%     \item \textbf{UFEF} \cite{ying2023unified}: A unified frontend framework to capture the dependency relationship between English TTS frontend modules, based on BERT as the skeleton, combined with rule-based methods.
% \end{itemize}

We compare our proposed TAP-FM with the following benchmark TTS frontend models:

\begin{itemize}
    \item \textbf{BERT-base} \cite{BERT}: Use pre-trained BERT as the text encoder and use MLP as the prediction layer for three subtasks.

    \item \textbf{SpeechT5} \cite{ao2021speecht5}: A character-level universal pre-training model that can complete TTS and ASR tasks. We integrate the text encoder into frontend modeling.

    \item \textbf{UFEF} \cite{ying2023unified}: A unified frontend framework to capture the dependency relationship between English frontend subtasks, based on BERT as the skeleton, combined with rule-based methods.
\end{itemize}

\begin{table*}[!t]
\caption{
Performance of our model and baselines on test sets.  The best performances are in \textbf{bold} and the second-best results are underlined. ``Improvement'' denotes the relative percentage improved by TAP-FM over the best baseline in each dataset.
}
\vspace{0.2cm}
\centering
\setlength{\tabcolsep}{0.45cm}
\begin{tabular}{ c|c c|c c c|c}

\specialrule{0.1em}{0em}{0em} % 在\hline上方添加粗线
\hline

\multirow{3}*{Method}    & \multicolumn{2}{c}{TN}  & \multicolumn{3}{|c|}{PBP} & PD  \\ 
\cline{2-7}

 & \multirow{2}*{F1-score (\%)} & \multirow{2}*{Acc (\%)} &  \multicolumn{3}{c|}{F1-score (\%)} &   \multirow{2}*{Acc (\%)} \\

\cline{4-6}
  &  & & PW& PPH & \multicolumn{1}{c|}{IPH} &  \\

\hline

BERT-based & 94.21 & 94.31 & \underline{95.17} & 62.67 & 76.33  & 95.19 \\

% \hline

SpeechT5 & 94.32 & 94.39 & 94.92 & 61.85 & 76.58 &  94.72 \\

% \hline

UFEF & \underline{94.89} & \underline{94.96} & 95.02 & \underline{62.88}  & \underline{76.91} &  \underline{95.68} \\

% \hline

TAP-FM & \textbf{96.32} & \textbf{96.18} &  \textbf{96.37} & \textbf{66.05} & \textbf{78.02} &  \textbf{97.08}  \\

\hline 

Improvement & 1.51\% & 1.28\% & 1.26\% & 5.04\% & 1.44\%   & 1.50\% \\

\specialrule{0.1em}{0em}{0em} % 在\hline上方添加粗线
\hline
\end{tabular}
\label{tab:result}
\end{table*}
\subsection{Comparisons with State of the Arts}

% 我们把每个实验运行三次，平均结果展示在表中，从中我们可以观察到以下结果。
% Among the three strong baseline models, BERT-based performs the worst in the TN task. Unlike SpeechT5, which relies on English characters for tokenization during pre-training, BERT-based is pre-trained with pure text and lacks audio features. UFEF, on the other hand, enhances local feature extraction with CNN, resulting in improved text modeling.

% SpeechT5 performs the worst in tasks related to predicting prosodic boundaries and resolving ambiguities in homophones due to its tokenization based on English characters. However, it performs slightly better in the TN task, where most NSWs are composed of special characters.

% UFEF outperforms all baseline models with a slight advantage. In the TN task, UFEF combines the advantages of rules and CRF in decoding prediction results. In the PBP task, UFEF treats the three rhythmic levels as three binary classification tasks, resulting in slight gains in PPH and IPH, although it lags behind BERT-based in PW. Additionally, UFEF uses the CBG module to enhance context modeling, leading to improved performance in the PD task.

% Our proposed TAP-FM achieves the best performance across all tasks. The largest improvement is 5.04\% in the PBP task, demonstrating the effectiveness and generality of TAP-FM in the TTS frontend.

As shown in Table \ref{tab:result}, our proposed TAP-FM achieves the best performance across all tasks, with the largest improvement being 5.04\% in the PBP task. This demonstrates the effectiveness and generality of TAP-FM on the TTS frontend.

Among the three baseline models, the BERT-based model performs slightly worse in the TN task. This is because SpeechT5 relies on English characters for tokenization during pre-training, while BERT-based is pre-trained with pure text and lacks audio features. On the other hand, UFEF enhances local feature extraction with CNN, leading to improved text modeling. Baseline SpeechT5 performs the worst in tasks PBP and resolving ambiguities in homophones due to its tokenization based on English characters. However, it performs slightly better in the TN task, where most NSWs are composed of special characters. Considering the UFEF model, it combines the advantages of rules and CRF in decoding prediction results in the TN task. In the PBP task, UFEF treats the three rhythmic levels as three binary classification tasks, resulting in slight gains in PPH and IPH, although it lags behind BERT-based in PW. Additionally, UFEF uses the CBG module to enhance context modeling, leading to improved performance in the PD task.

\subsection{Analysis on Text-Audio Alignment}
% \subsection{Quantitative Evaluation}
\begin{table}
\caption{Average text-audio cosine similarity of BERT and MC-TAP, respectively. }
\vspace{0.2cm}
\centering
\setlength{\tabcolsep}{0.4cm}
\begin{tabular}{c|c|c}
\specialrule{0.1em}{0em}{0em} % 在\hline上方添加粗线
    \hline
    Method & BERT& MC-TAP \\
    \hline
     Average similarity& -0.0005 & 0.8123 \\
     \specialrule{0.1em}{0em}{0em} % 在\hline上方添加粗线
    \hline
\end{tabular}

\label{tab:avg_sim}
\end{table}
\begin{table*}[!t]
\caption{
Extensive studies to ablate the impacts of our proposed individual components.
}
\vspace{0.2cm}
\centering
\setlength{\tabcolsep}{0.25cm}
\begin{tabular}{  c | l|c  c| c c c|c}
\specialrule{0.1em}{0em}{0em} % 在\hline上方添加粗线
\hline
 \multicolumn{2}{c|}{}     & \multicolumn{2}{c}{TN}  & \multicolumn{3}{|c|}{PBP} & PD  \\ 
\cline{3-8}

\multicolumn{2}{c|}{Method}  & \multirow{2}*{F1-score (\%)} & \multirow{2}*{Acc (\%)} &  \multicolumn{3}{c|}{F1-score (\%)} &   \multirow{2}*{Acc (\%)} \\

\cline{5-7}
 \multicolumn{2}{c|}{}  & & & PW& PPH & IPH  & \\
 
\hline
\multicolumn{2}{c|}{TAP-FM} & 96.32 & 96.18 &  \textbf{96.37} & \textbf{66.05} & \textbf{78.02} & \textbf{97.08} \\
\hline
 & w/o Span-level  & 95.46 & 95.56 & 95.32 & 64.55 & 77.93 &  96.21  \\ 

% \cline{2-8}
{\makecell{Stage one}} & w/o Sentence-level & 96.07 & 95.89 & 96.02 & 65.75 & 77.95 &  96.88 \\

% \cline{2-8}
& w/o MLM & 95.67 & 95.36 & 95.79 & 65.03 & 77.38 & 96.33 \\

% \cline{2-8}
% &— w/o Cross & & & & & &  \\

\hline
& w/o TE from MC-TAP & 94.96 & 95.08 & 95.32 & 63.85 & 77.25 &  95.78  \\

{\makecell{Stage two}}& w/o ResConformer 
& 95.78 & 95.96 & 95.12 & 64.81 & 77.39 &  96.56 \\

% \cline{2-8}
& w/o DWA+ 
& \textbf{96.38} & \textbf{96.41} & 95.16 & 64.93 & 76.98 &  96.16\\
\specialrule{0.1em}{0em}{0em} % 在\hline上方添加粗线
\hline
\end{tabular}
\label{tab:ablation}
\end{table*}
% 我们在Kaldi工具的辅助下，人工提取了每个单词对应的音频片段
\noindent \textbf{Quantitative Evaluation.} We sampled 2.7k text-audio pairs from the validation set during the pre-training stage, including 54k words.
We have manually extracted audio segments corresponding to each word with the help of the Kaldi\footnote{\url{https://eleanorchodroff.com/tutorial/kaldi}} tool,
and then obtain audio representations through the AudioEncoder. For each word, we use both BERT and MC-TAP to obtain word vector representations and calculate their cosine similarity with the audio representation. The average similarity is shown in Table \ref{tab:avg_sim}.
BERT's text-audio similarity is only -0.0005. We can infer that text-based pre-trained language models do not have audio features, which limits the performance of these models in the TTS field. After applying multi-scale contrastive text-audio pre-training, the similarity between the word vectors obtained by MC-TAP and audio features significantly improves to 0.8123. This indicates that our pre-training strategy is effective and achieves the expected results.

% \subsection{Case Study}\label{sec:case_study}
\begin{figure}
    \centering
    \includegraphics[width=1\linewidth, trim=10 0 10 0, clip]{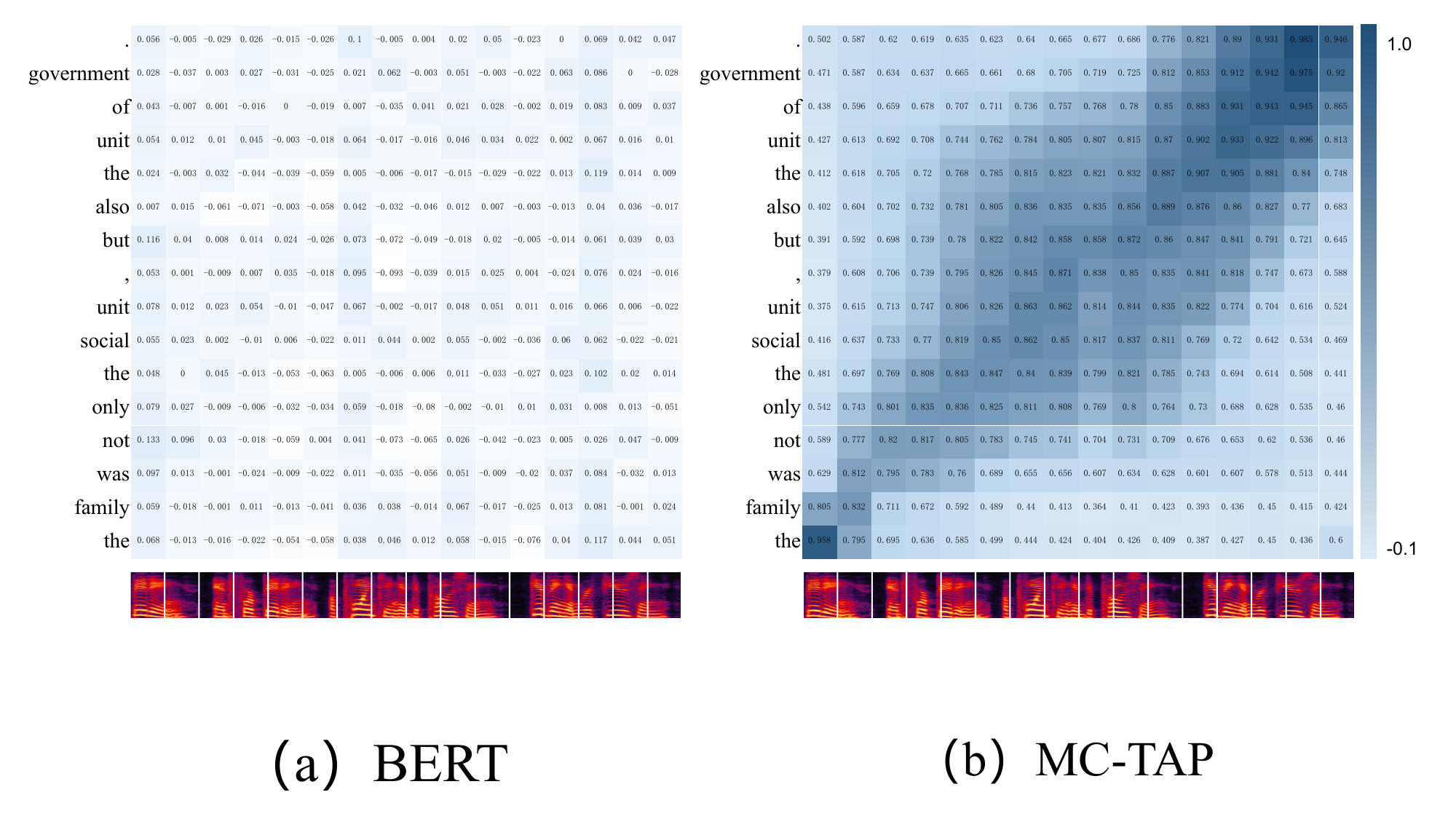}
    % 词向量与对应音频片段特征之间的余弦相似度矩阵，词向量分别来自于BERT和MC-TAP，颜色越深代表相似度越高。
    \caption{The cosine similarity matrix between word vectors and corresponding audio segment features from a real case. Darker colors represent higher similarity.}
    \label{fig:sim-hot}
\end{figure}
% 我们从验证集中选取了内容为``你好''的一个句子，并可视化了由BERT和MC-TAP生成的词向量与对应单词音频特征的余弦相似度矩阵，结果展示在图中。

\vspace{0.2cm}
\noindent \textbf{Case Study.} We selected a sentence with the content ``\textit{The family was not only the social unit, but also the unit of government.}'' from the validation set and visualized the cosine similarity matrix between the word vectors generated by BERT and MC-TAP and the corresponding word audio features. The result is shown in Figure \ref{fig:sim-hot}.
% BERT的词向量跟对应的音频表征之间的整体相似度较低，而且没有规律可循。MC-TAP的相似度矩阵呈现明显的对角相似性，平均相似度较高。另外，同一个单词在句子中跟非对应位置的声音片段表现出较弱的相关性，比如“the”和“unit”，这表明MC-TAP建模的信息是强上下文相关的。

The overall similarity between BERT's word vectors and their corresponding audio representations is relatively low and lacks a discernible pattern. In contrast, MC-TAP's similarity matrix exhibits clear diagonal similarities and a higher average similarity. Furthermore, the same word in a sentence shows a weaker correlation with non-corresponding audio segments, such as ``the'' and ``unit'', indicating that the information modeled by MC-TAP is highly context-dependent.

\subsection{Ablation Study}

% TAP-FM is developed with a multi-scale contrastive learning strategy during the pre-training stage, incorporating the ResConformer structure and the DWA+ multi-task joint learning optimization method in downstream tasks. 
To assess the effectiveness of the components from TAP-FM, we examined various variants of TAP-FM:
\emph{(1)} ``\textbf{w/o Span-level}'' refers to the exclusion of span-level contrastive learning during the pre-training phase.
\emph{(2)} ``\textbf{w/o Sentence-level}'' implies the absence of positive and negative sample sampling for entire sentences during pre-training as part of the contrastive learning process.
\emph{(3)} ``\textbf{w/o MLM}'' signifies the omission of the Masked Language Model task during pre-training.
\emph{(4)} ``\textbf{w/o TE from MC-TAP}'' denotes the non-utilization of TextEncoder from MC-TAP, instead of initializing model parameters solely from origin BERT.
\emph{(5)} ``\textbf{w/o ResConformer}'' indicates the exclusion of the Conformer as a residual structure in downstream tasks.
\emph{(6)} ``\textbf{w/o DWA+}'' implies the absence of the DWA+ strategy during multi-task joint training.
The experimental results comparing TAP-FM and its variants are presented in Table \ref{tab:ablation}. Our observations yield the following insights:

\noindent$\bullet$ \textbf{Impact of Span-level Contrastive Learning.} During the pre-training phase, the most significant performance drop occurs when we remove the Span-level alignment task. This highlights the critical role of close-range dependencies within audio data for downstream frontend modeling. Fine-grained contrastive learning facilitates the alignment of text and audio at the hidden layer level.

\noindent$\bullet$ \textbf{Impact of Sentence-level Contrastive Learning.}
In contrast, eliminating Sentence-level contrastive learning has a relatively minor impact, suggesting that local information holds greater importance than global information in TTS tasks.

\noindent$\bullet$ \textbf{Impact of MLM.}
The exclusion of the MLM task also affects model performance to some extent. TTS tasks demand both semantic and acoustic features, and the loss of acoustic information during contrastive learning leads to performance degradation.

\noindent$\bullet$ \textbf{The ability of TE from MC-TAP.}
Using BERT to replace the TextEncoder form MC-TAP results in a significant performance decrease, as the model lacks crucial audio-related information. 
% 虽然TN任务跟声学的相关性较弱，但是预训练后的标准词的特征和NSW的特征会出现明显差距，会对TN任务带来收益
Although the correlation between TN and acoustics is weak, there will be a significant difference in the features of standard words after pre-training and the features of NSWs, which will bring benefits to the TN task.

\noindent$\bullet$ \textbf{Effect of ResConformer.}
In the fine-tuning of downstream tasks, the removal of the residual structure composed of the Conformer network leads to faster coverage of learned audio features, causing a decline in model performance.

\noindent$\bullet$ \textbf{Effect of DWA+.}
Additionally, when conducting multi-task joint training without employing the DWA+ strategy, model parameter updates predominantly focus on the TN task, resulting in insufficient training for the PBP and PD tasks, primarily due to the TN problem's higher complexity.

%% file: sections/7_conclusion.tex
\section{Conclusion}
% To tackle this issue, a novel two-stage TTS frontend prediction pipeline is introduced. In the first stage, we develop an unsupervised Multi-scale Contrastive Text-audio Pre-training (MC-TAP) framework,  which utilizes multi-grain contrastive pre-training to learn more abundant insights.  Compared to existing pre-training methods, this framework can capture both global and local text-audio semantic representation to improve downstream prediction tasks. Additionally, our approach eliminates the need for preprocessing data by using automatic alignment between text and audio, thereby reducing a significant amount of preprocessing costs.  In the following stage,  we design a fully parallelized TTS frontend model for predicting TN, PD and PBP tasks. The model employs the pre-trained text encoder to extract refined text-audio semantics. Our experiments demonstrate that our proposed system can achieve state-of-the-art (SOTA) performance for TTS frontend prediction.

In this paper, we propose a two-stage method called TAP-FM, which aims to improve the accuracy and expression of the TTS frontend tasks. First,  we develop a prior-agnostic multi-scale contrastive text-audio pre-training model to learn the automatic alignment of semantic and acoustic information on text and audio pairs. Second, we propose a parallelized model for predicting TN, PBP, and PD tasks. 
The extensive experiments clearly illustrate the superiority of the TAP-FM model that has been put forward.